\documentclass[journal=jacsat,manuscript=article]{achemso}

\usepackage{ulem}
\usepackage{times}
\usepackage[version=3]{mhchem} 
\usepackage[colorlinks,linkcolor=blue,citecolor=blue,urlcolor=black]{hyperref}

\author{Rong Hu}
\author{Yu-Tao Tan}
\author{Dapeng Liu}
\author{Jie Ren}
\email{xonics@tongji.edu.cn}
\author{Yizhou Liu}
\email{yizhouliu@tongji.edu.cn}
\affiliation{Center for Phononics and Thermal Energy Science, China-EU Joint Lab on Nanophononics, School of Physics Science and Engineering, Tongji University, 200092 Shanghai, China}

\title{Moir\'e Strain Skyrmions in Sliding Twisted Bilayers}

\begin{document}
\begin{tocentry}

\includegraphics[width=\linewidth]{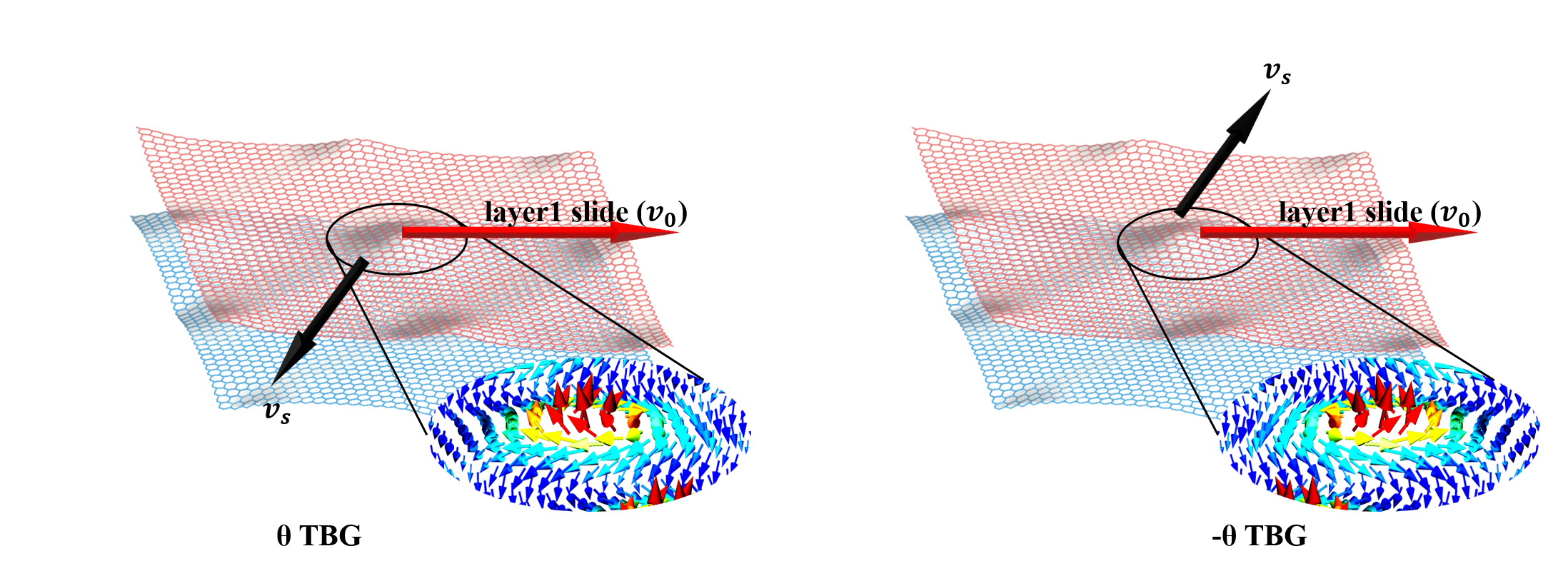}





\end{tocentry}

\begin{abstract}
Strain defect is crucial to the physical properties of solid materials. Among them, strain glass induced by defect engineering provides an important paradigm for nanoscale domain manipulation. Here, we propose purely mechanical moir\'e strain Skyrmions, a topologically protected elastic textures whose motion can be controlled by interlayer sliding and the chirality of the moir\'e bilayer. Using an empirical continuum elastic model combined with symmetry analysis, we demonstrate the Skyrmion lattice structure as the elastic ground state. Under interlayer sliding, these moir\'e strain Skyrmions exhibit the Skyrmion Hall effect of transverse motion, with a Hall angle determined by bilayer chirality and inversely proportional to the moir\'e twist angle. Our work establishes interlayer sliding as an efficient, low-energy control knob for topological excitations, offering a new paradigm for designing chiral-material-based information transport devices.

\end{abstract}

\maketitle

 




\section{Introduction}
The strain defect serves as a powerful inducer for tuning the physical, chemical, and electronic properties of solid-state materials \cite{nelson2002defects, RevModPhys.81.109, hou2025strain, dai2019strain, feng2012strain, peng2020strain, si2016strain}. In crystalline solids, strain defects arise from a range of sources, such as dislocations \cite{yazyev2010topological, Stukowski_2010, hull2011introduction}, point defects\cite{RevModPhys.86.253, Macdonald_1992, MACDONALD20111761}, grain boundaries \cite{wei2012nature, yazyev2010topological} and lattice mismatches \cite{KOMA1999236, doi.10.1021}. Strain fields are a type of lattice defects, in two-dimensional(2D) materials, strain defects acquire particular significance due to its abnormal properties \cite{si2016strain}, such as altering electron-phonon coupling \cite{PhysRevLett.111.196802}, reshaping energy band \cite{feng2012strain,si2016strain}, superconductivity\cite{PhysRevLett.111.196802} and so on. Consequently, the ability to manipulate strain defects through external stimulation affords precise control over material functionalities, making strain defects a cornerstone of modern materials
design for smart and adaptive systems.

In recent years, the discovery of several exotic strain-induced phase transition phenomena has fundamentally overturned our understanding of phase of elastic matter. Notably, the strain glass state—a disordered state of frozen martensitic nanodomains—was first reported in 2005 by Sarkar et al \cite{sarkar2005evidence} and has since been extensively investigated \cite{wang2010strain,wang2006shape}. This unique state exhibits many intriguing physical properties, including high mechanical strength \cite{xu2024polymer}, large piezoelectric responses \cite{liu2009large}, switchable ferroelectricity \cite{ren2004large}, and a colder-is-faster transition behavior that contrasts with traditional oxide and metallic glasses\cite{zhang2016accelerating}. 
The strain glass paradigm has broadened the theory of phase transformations in ferroelastic systems, which provides new avenues for nanodomain engineering. However, artificially engineering such as phase necessitates a platform where long-range periodic frustration naturally replaces the role of random defects in stabilizing these nanoscale domains.  In parallel, the rapid development of topological phononics has revealed that lattice vibrations and elastic deformations can also exhibit rich topological properties, including spin in acoustics\cite{long2018intrinsic,shi2019observation,long2020realization,long2020symmetry,yuan2021observation,ren2022elastic,yang2023hybrid,yang2024chirality},phonon edge state\cite{liu2017pseudospins,PhysRevLett.108.086602} and chiral phonon transport\cite{liu2018berry,liu2020topological,liu2022ubiquitous}. In particular, recent advances in quantum phononics have demonstrated that phonon spin and chirality can serve as new degrees of freedom for manipulating mechanical information, opening up exciting possibilities for chiral mechanical devices\cite{lei2025quantum,zhao2022elastic,liu2021chirality}. \par
\begin{figure}
    \centering
    \includegraphics[width=\linewidth]{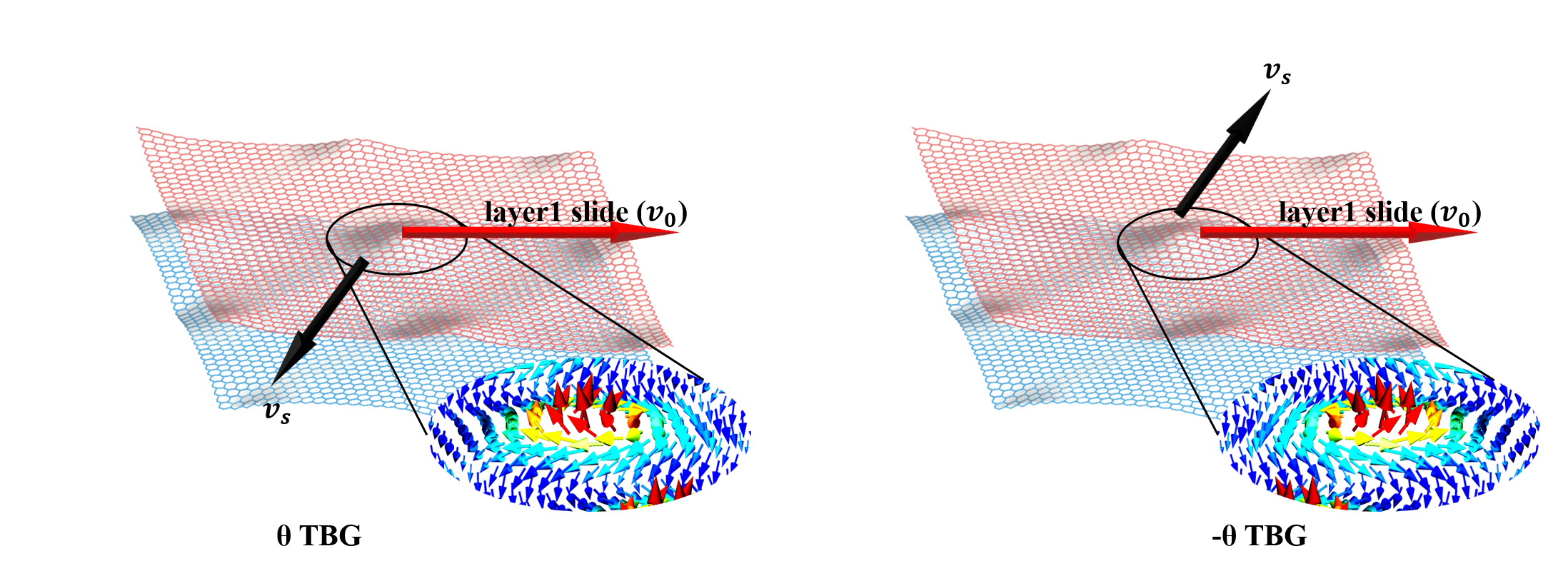}
    \caption{Strain Skyrmion Hall effect induced by interlayer sliding in twisted bilayer graphene (TBG). Left panel: Under a counter-clockwise twist angle ($\theta$), the drift velocity of the skyrmion ($\textbf{v}_s$) is proportional to the sliding velocity ($\textbf{v}_0$), showing a positive linear Hall conductivity, i.e. $v^x_s = \sigma_{xy} v^y_0$ with $\sigma_{xy}>0$. Right panel: Under a clockwise twist angle ($-\theta$), the Hall conductivity becomes negative.}
    \label{fig:1}
\end{figure}

In this work, we propose a controllable topological strain phase, the strain Skyrmion lattice, in twisted bilayer graphene (TBG), wherein the Skyrmion lattice sites are highly tunable via interlayer sliding as shown in Fig. \ref{fig:1}. Based on an
continuum elastic model incorporating both in-plane and out-of-plane degrees of freedom, we find that the ground-state displacement field of each layer, with respect to the untwisted case, exhibits a topological structure characterized by quantized Skyrmion number within each moir\'e supercell thus forming a strain Skyrmion triangular lattice. Furthermore, we demonstrate that by introducing interlayer sliding adiabatically, the strain Skyrmions exhibit a topological Skyrmion Hall effect with the Hall angle (i.e. the ratio between the Skyrmion drift speed $v_s$ and the interlayer sliding speed $v_0$) controlled by the twist angle of the moir\'e lattice. We further generalize the strain Skyrmion Hall effect to arbitrary moir\'e systems with point group symmetries $C_n$ ($n=2,3,4,6$). 
This work establishes a theoretical foundation for the mechanical manipulation of topological strain quasiparticles, opening new avenues for designing chiral material devices and high-speed, mechanically driven information transport architectures. \par

\section{Strain Skyrmion in Ground State of TBG}
\begin{figure*}[t]
    \centering
    \includegraphics[width=1\linewidth]{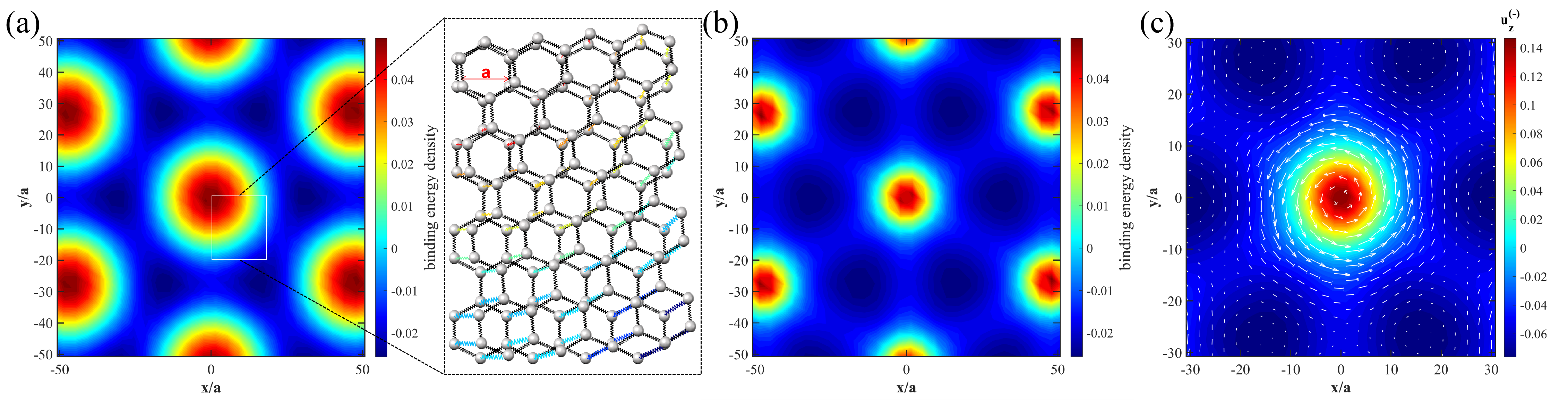}
    \caption{(a) Real-space distribution of interlayer binding energy $U_B$ before relaxation in twisted bilayer graphene (TBG). The right panel shows a zoom-out real-space atomic structure of TBG. The red (blue) colored springs represent higher (lower) interlayer binding energy. (b) Real-space distribution of $U_B$ after relaxation in TBG. (c) The relaxation displacement field of the TBG, with white arrows indicating in-plane components ($u^{(-)}_{x,y} = u^{(2)}_{x,y} - u^{(1)}_{x,y}$) and the background color represents the out-of-plane component ($u^{(-)}_z = u^{(2)}_z - u^{(1)}_z$). $\textbf{u}^{(-)} = \textbf{u}^{(2)} - \textbf{u}^{(1)}$ represents the interlayer relative displacement field vector. This displacement field can be characterized by a topological Skyrmion number $N = 1$. }
    \label{fig:2}
\end{figure*}
We consider a TBG lattice with a relatively small twist angle $\theta$, typically a few degrees. Under the condition, the wavelength of the moir\'e system is far larger than the lattice constant, justifying the use of a continuous elastic model. We introduce the displacement field $\textbf{u}^{(l)}= (u^{(l)}_x, u^{(l)}_y, u^{(l)}_z)$ to describe in-plane and out-of-plane displacement terms of $l$ th layer ($l\in\{1,2\}$). We further define the symmetric and antisymmetric displacement combinations as $\textbf{u}^{(\pm)}=\textbf{u}^{(2)}\pm \textbf{u}^{(1)}$. The total elastic energy of the TBG system can be decomposed into an intralayer term and an interlayer binding energy term. The intralayer term comprises a standard elastic term, expressed as $U_E = \sum^2_{l=1} \sum_{\alpha,\beta=x,y} \int \mathrm{d}^2\textbf{r}~ [ \frac{\lambda}{2}(\partial_{\alpha} u_{\alpha}^{(l)})^2 + \frac{\mu}{4}(\partial _{\alpha }u_{\beta}^{(l)}+\partial _{\beta }u_{\alpha}^{(l)})^2] = \sum_{\alpha,\beta=x,y} \int \mathrm{d}^2\textbf{r}~ \frac{\lambda}{4}[ (\partial_\alpha u^{(+)}_\alpha)^2 + (\partial_\alpha u^{(-)}_\alpha)^2] + \frac{\mu}{8} [(\partial_\alpha u^{(+)}_\beta + \partial_\beta u^{(+)}_\alpha)^2 + (\partial_\alpha u^{(-)}_\beta + \partial_\beta u^{(-)}_\alpha)^2]$, where $\mu \approx 9.57 \, \text{eV\AA}^{-2}$ and $\lambda \approx 3.25 \, \text{eV\AA}^{-2}$ are the Lam\'e coefficients of the monolayer graphene\cite{koshino2019moir}. The interlayer binding energy ($U_B$) consists of an in-plane term ($U^\parallel_B$) and an out-of-plane term ($U^\perp_B$), i.e. $U_B = U^\parallel_B + U^\perp_B$, where $U^\parallel_B$ is expressed as $U^\parallel_B = \sum^3_{j=1} \int V_0 \cos[\textbf{G}_j^M \cdot \textbf{r} +\textbf{a}_j^* \cdot \textbf{u}^{(-)}]~ \mathrm{d}^2 \textbf{r}$\cite{nam2017lattice,koshino2019moir,suri2021chiral}. $\textbf{a}^*_j$ ($j=1,2,3$) are the reciprocal lattice vectors corresponding to primitive graphene: $\textbf{a}^*_1 = \frac{4\pi}{\sqrt{3}a}(\frac{\sqrt{3}}{2},-\frac{1}{2},0)$, $\textbf{a}^*_2 = \frac{4\pi}{\sqrt{3}a} (0,1,0)$, and $\textbf{a}^*_3=-\textbf{a}^*_1 - \textbf{a}^*_2$ ($a$ is the lattice constant); $\textbf{G}^M_j$ are the corresponding reciprocal lattice vectors of moir\'e supercell: $\textbf{G}^M_j = [1 - \mathsf{R}_z(-\theta)] \textbf{a}^*_j$ ($\mathsf{R}_z(.)$ refers to the rotation operation along $z$ axis). The out-of-plane term is expressed as 
\begin{equation}
U^{\perp}_B = \int d^2 \textbf{r}~ C [u_z^{(-)} + \bar{d} - d_{\textrm{opt}}]^2,
\end{equation}
where $C$ is a constant, $\bar{d}$ refers to the interlayer distance of untwisted bilayer graphene, and $d_{\textrm{opt}}$ is the optimized interlayer distance in the presence of interlayer twist. We assume that the optimized interlayer distance is given by
\begin{equation}
    d_{\textrm{opt}} = \bar{d} + \sum^3_{j=1} \frac{2}{9} \Delta d \cos[\textbf{G}^M_j \cdot \textbf{r} + \textbf{a}^*_j \cdot \textbf{u}^{(-)}],
\end{equation}
where the maximum value of $d_{\textrm{opt}}$ is $d^{\textrm{max}}_{\textrm{opt}} =\bar{d} + \frac{2}{3}\Delta d$ and the minimum value is $d^{\textrm{min}}_{\textrm{opt}} = \bar{d} - \frac{1}{3}\Delta d$. $\Delta d = |d^{\textrm{max}}_{\textrm{opt}} - d^{\textrm{min}}_{\textrm{opt}}|$ is the maximum interlayer distance fluctuation which is easily obtained by first-principles calculations. Maximized interlayer distance is $d^{\textrm{max}}_{\textrm{opt}} = 3.60$ \AA~with AA stacking and $d^{\textrm{min}}_{\textrm{opt}} = 3.35$ \AA~with AB stacking\cite{lin2018shear}.

Next, we determine the elastic ground state of TBG by minimizing the elastic potential energy. The minimized elastic potential energy can be determined by $\delta (U_E + U^\parallel_B + U^\perp_B)/\delta \textbf{u}^{(+)} = 0$ and $\delta (U_E + U^\parallel_B + U^\perp_B)/\delta \textbf{u}^{(-)} = 0$. Since the total potential energy is a semi-positive-definite function of $\textbf{u}^{(+)}$, the ground-state energy $U_E + U^\parallel_B + U^\perp_B$ corresponds to $\textbf{u}^{(+)}=0$ and the total potential energy landscape only relies on $\textbf{u}^{(-)}$. Figures \ref{fig:2}(a)-(b) illustrates the potential energy landscapes before and after twisting, where warmer (colder) colors indicate higher (lower) local stacking energies. It is clearly observed that before and after twisting, the system undergoes spontaneous lattice relaxation, which can be treated as an adiabatic evolution process relative to the macroscopic system dynamics. Specifically, the AA stacking with high energy undergoes significant spatial shrink, whereas the AB/BA stacking with low energy undergoes significant spatial expansion,  the corresponding displacement field of this process has been shown in Fig. \ref{fig:2}(c). The complete 3D displacement field, encompassing both in-plane and out-of-plane relaxations, supports a topological Skyrmion number defined as
\begin{equation}
    N = \frac{1}{4\pi} \iint_{\textrm{u.c.}} \textbf{n} \cdot \left( \frac{\partial \textbf{n}}{\partial x} \times \frac{\partial \textbf{n}}{\partial y} \right)~ \mathrm{d}x\mathrm{d}y,
\end{equation}
where $\textbf{n} = \textbf{u}^{(-)}/|\textbf{u}^{(-)}|$ and spatial integration is restricted within one moir\'e unit cell. Since the unit direction field $\textbf{n}(\textbf{r})$ of each strain skyrmion is a localized, asymptotically uniform texture at boundary, the two-dimensional plane can be one-point compactified to a sphere $S^2$. As such, the skyrmion number is classified by the second homotopy group $\pi_ 2(S^2)=Z$, making it a quantized and topologically protected charge. This confirms the moir\'e strain texture is indeed a  topological strain skyrmion in real space.

Our numerical calculation shows $N=+1$ for both clockwise and counter-clockwise twisting angles $\theta$, but the resulting Skyrmions have different helicities depending on the sign of $\theta$, i.e. chirality of the TBG. The topological number $N$ does not rely on the values of Lam\'e coefficients $\lambda$ and $\mu$ indicating interlayer twisting as a general way for emergent topological strain textures.

\begin{figure*}
    \centering
    \includegraphics[width=\linewidth]{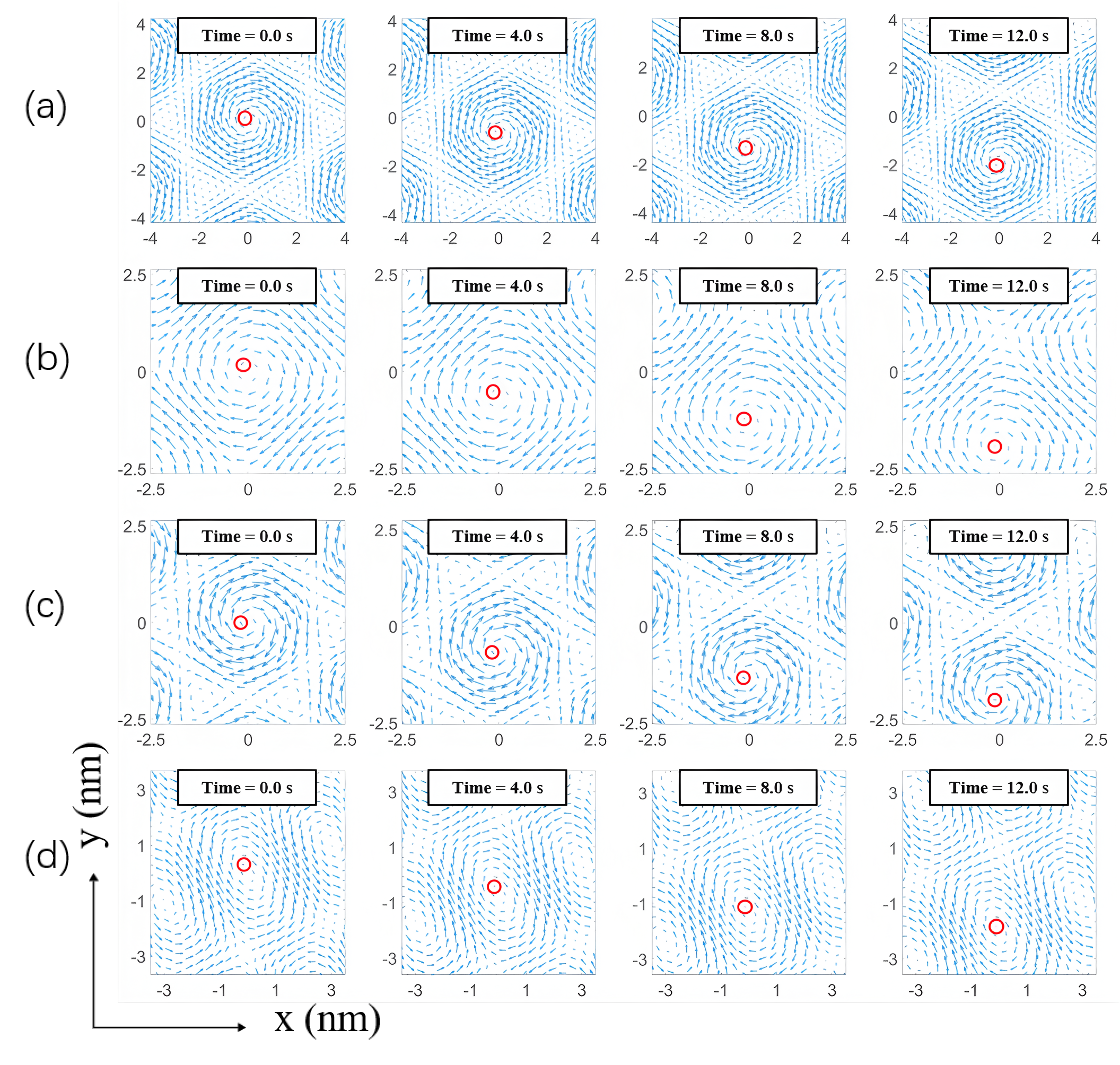}
    \caption{Vortex motion induced by inter-layer sliding in moir\'e bilayers with different point group symmetries. (a) Motion of vortex induced by sliding with $C_6$ symmetry. (b)-(d) Same as (a) but with point group symmetries $C_4$, $C_3$, and $C_2$, respectively. The elastic parameters $\lambda$, $\mu$, $V_0$ are taken as same as TBG. The red circles indicate the vortex centers. }
    \label{fig:3}
\end{figure*}

\section{Strain Skyrmion Hall Effect}

We investigate the dynamical properties of moir\'e strain Skyrmions in TBG under continuous interlayer sliding. The interlayer sliding can be simulated by replacing $\textbf{u}^{(-)}$ by $\textbf{u}^{(-)} + \textbf{v}_0t$ in the binding energy terms $U^\parallel_B$ and $U^\perp_B$, with $\textbf{v}_0$ representing the interlayer sliding velocity, i.e.
\begin{equation}
    \begin{split}
        U^\parallel_B \rightarrow \tilde{U}^\parallel_B =& \int d^2 \textbf{r} \sum^3_{j=1} V_0 \cos[ \textbf{G}^M_j \cdot \textbf{r} + \textbf{a}^*_j \cdot (\textbf{u}^{(-)} + \textbf{v}_0 t) ], \\
        U^\perp_B \rightarrow \tilde{U}^\perp_B =& \int d^2\textbf{r}~ C[ u^{(-)}_z + \bar{d} - \tilde{d}_{\textrm{opt}} ]^2, \\
        \tilde{d}_{\textrm{opt}} =& \bar{d} + \sum^3_{j=1} \frac{2}{9} \Delta d \cos [ \textbf{G}^M_j \cdot \textbf{r} + \textbf{a}^*_j \cdot ( \textbf{u}^{(-)} + \textbf{v}_0 t ) ]. 
    \end{split}
\end{equation}

For a sufficiently small value of $\textbf{v}_0$, the effect of this relative displacement can be approximately treated as an adiabatic process. For a horizontal interlayer sliding $\textbf{v}_0 = v_0 \hat{\textbf{e}}_x$ with $v_0 = 0.1$ \,\AA/s, the time evolution of the ground-state distribution of $\textbf{u}^{(-)}$ determined by $\delta(U_E + \tilde{U}^\parallel_B + \tilde{U}^\perp_B)/\delta\textbf{u}^{(-)} = 0$ is shown in Fig. \ref{fig:3}(a), more examples can be seen in Supplemental Material\cite{my_supplementary}. We find that at a generic time $t$ the distribution of $\textbf{u}^{(-)}$ is still a strain Skyrmion with $N=+1$. As $t$ increases, the strain Skyrmion centers move vertically with constant speed $v_s$ indicating an interlayer sliding driven strain Skyrmion Hall effect. 

Such a sliding driven strain Skyrmion Hall effect generally exist in two-dimensional moir\'e systems with point group symmetries $C_n$ ($n=2,3,4,6$). The binding energy density $\mathcal{U}^\parallel_B$, $\mathcal{U}^\perp_B$ is constructed as
\begin{equation}
    \begin{split}\label{binding_energy_density_Cn}
        \mathcal{U}^\parallel_B =& \int d^2 \textbf{r} \sum^n_{j=1} V_0  \cos [ \textbf{G}^M \cdot \textbf{r}' + \textbf{a}^* \cdot ( \textbf{u}^{(-)'} + \textbf{v}_0 t) ], \\
        \mathcal{U}^\perp_B =& \int d^2 \textbf{r} C[u^{(-)}_z + \bar{d} - \tilde{d}'_{\textrm{opt}}], \\
        \tilde{d}'_{\textrm{opt}} =& \bar{d}+ \sum^n_{j=1} \frac{2}{9} \Delta d  \cos[\textbf{G}^M \cdot \textbf{r}' + \textbf{a}^* \cdot (\textbf{u}^{(-)'} + \textbf{v}_0 t)],
    \end{split}
\end{equation}
where $\textbf{r}'=C_n^{-j}\textbf{r}$ and $\textbf{u}^{(-)'} = C^j_n \textbf{u}^{(-)}(C^{-j}_n \textbf{r})$ represent the rotation transformation of $\textbf{r}$ and $\textbf{u}^{(-)}(\textbf{r})$, respectively. $C^j_n = \mathsf{R}_z(2\pi j/n)$ is the rotation symmetry operation. $\textbf{a}^*$ and $\textbf{G}^M$ represents one of the reciprocal Bravais lattice vector for the primitive monolayer and its corresponding moir\'e bilayer. Figures \ref{fig:3}(b)-(d) show the time evolution of strain vortex for $n=4,3,2$, respectively. Two conclusions can be drawn: (i) The strain Skyrmion is the ground state of moir\'e system with arbitrary rotational symmetry $C_n$; (ii) In the presence of interlayer sliding, the strain Skyrmion lattices generally exhibit Hall effect with similar Hall drift velocity. 

In order to quantitatively describe the strain Skyrmion Hall effect, we rewrite the argument in the first and third lines of Eq. \eqref{binding_energy_density_Cn} as
\begin{equation}
    \begin{split}\label{replacement}
        \textbf{G}^M \cdot \textbf{r} + \textbf{a}^* \cdot (\textbf{u}^{(-)} + \textbf{v}_0 t) = \textbf{G}^M \cdot (\textbf{r} - \textbf{v}_s t) + \textbf{a}^* \cdot \textbf{u}^{(-)}.
    \end{split}
\end{equation}
The above equation indicates that the relative interlayer sliding $\textbf{v}_0t$ can be equivalent to the movement of strain field $\textbf{u}^{(-)}(\textbf{r})$ by replacing $\textbf{r}$ by $\textbf{r} - \textbf{v}_s t$. $\textbf{v}_s$ is the moving velocity of the strain field. Combining Eq. \eqref{replacement} with the relation $\textbf{G}^M_j = [1-\mathsf{R}_z(-\theta)]\textbf{a}^*_j$, we can derive that
\begin{equation}
    \begin{split}\label{vs}
        \textbf{v}_s =& -\{[ 1- \mathsf{R}_z(-\theta)]^\top\}^{-1} \textbf{v}_0 = \frac{1}{\theta} \left(
        \begin{array}{cc}
            0  & 1 \\
            -1 & 0 
        \end{array}
        \right) \textbf{v}_0 + O(\theta^2).
    \end{split}
\end{equation}
For small value of $\theta$, the Skyrmion velocity $\textbf{v}_s$ is proportional to sliding velocity $\textbf{v}_0$. We define the Skyrmion Hall angle $\theta_{\textrm{H}}$ as $\theta_{\textrm{H}} \equiv |\textbf{v}_s|/|\textbf{v}_0| = 1/\theta$. That is, the smaller twist angle $\theta$ is the larger the Skyrmion Hall angle $\theta_{\textrm{H}}$ is. It is interesting to note that the Skyrmion Hall angle only depends on the twist angle $\theta$ but does not rely on other details of the moir\'e lattice.

\begin{figure}
    \centering
    \includegraphics[width=\linewidth]{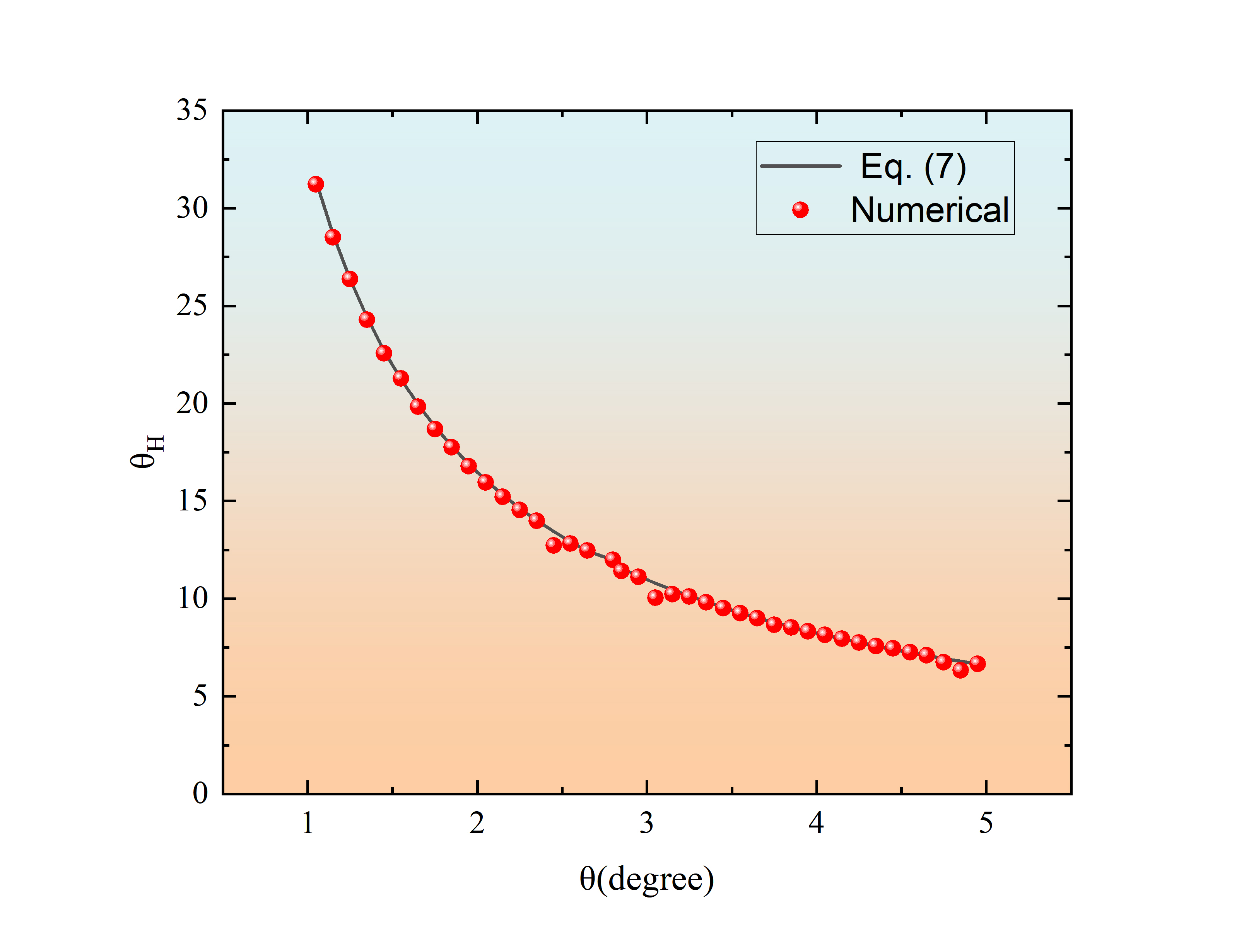}
    \caption{Comparison of theoretical values and numerical simulation results of the Skyrmion Hall angle $\theta_{\textrm{H}} = v_s/v_0$ as a function of twisting angle $\theta$.}
    \label{fig:4}
\end{figure}

To quantitatively verify the accuracy of the numerical simulation calculations, we compared the Skyrmion Hall angle $\theta_{\textrm{H}}$ with various twist angle $\theta$, and compared these numerical results with the theoretical velocity values derived from Equation (7). The results are shown in Fig. \ref{fig:4} and exhibit a remarkable high goodness of fit, confirming that our numerical simulations perfectly capture the theoretical topological kinematics within the margin of error. 

Our results demonstrate that moir\'e systems across all tested symmetries exhibit this linear topological Hall effect, underscoring the universality of the kinematic law. Fundamentally, this process can be mapped onto a mechanical Thouless pumping mechanism\cite{thouless1983quantization,zhang2020topological}, 
because after completing a full spatial period of interlayer sliding, the topological skyrmions are discretely pumped by a precisely quantized transverse displacement. 

\section{Discussion}
The fundamental upper bound of $v_s$ is limited by the lattice relaxation time, or the average phonon lifetime $\tau_{\textrm{ph}}$, which is typically at the order of picoseconds. Generally, a smaller $\tau_{\textrm{ph}}$ allows a faster relaxation of the strained lattice and thus a higher upper bound of $v_s$. Take $\tau_{\textrm{ph}} = 1$\,ps\cite{nika2017phonons,bonini2012acoustic,chen2012thermal}, the adiabatic condition requires that $v_s \ll \mathrm{max}\{|\textbf{u}^{(-)}|\}/\tau_{\textrm{ph}} \approx 32$ m/s, where $\mathrm{max}\{|\textbf{u}^{(-)}|\}=0.32$\,\AA~ is the maximum value of $|\textbf{u}^{(-)}(\textbf{r})|$ in magic angle TBG.
Furthermore, the interlayer sliding can further reduce the phonon lifetime $\tau_{\textrm{ph}}$ due to interlayer friction \cite{koren2016moire,PhysRevMaterials.9.024002,koren2015adhesion} which favors an even larger upper bound of $v_s$ and Skyrmion Hall angle $\theta_{\textrm{H}}$. 
Given that phonon relaxation timescales are typically much shorter than the timescales of macroscopic mechanical manipulation, this dynamical limit warrants further experimental investigation\cite{koren2016moire,hou2025strain,PhysRevMaterials.9.024002} and phonon dynamical evolution\cite{wang2024thermodynamic,lei2025quantum}.

In terms of the feasibility of the experiment, the purely mechanical origin of moiré strain skyrmions enables straightforward observation and manipulation with state-of-the-art techniques. 
Scanning tunneling microscope (STM) can map sub-angstrom atomic displacements, which can directly resolve the 3D strain skyrmion lattice, while second-harmonic generation (SHG) microscopy enables fast large-area detection via strain-gradient-induced inversion symmetry breaking\cite{yang2020tunable}. Also, interlayer sliding can be achieved by contact-mode AFM tips\cite{jiang2018manipulation,koren2015adhesion}, allowing deterministic control of individual skyrmions. Applying AFM to track skyrmion center trajectories in real time can directly quantify the Hall angle. Notably, unlike magnetic skyrmions which requires magnetic fields or currents with Joule heating, electronic skyrmions which requires cryogenic temperatures, and ferroelectric skyrmions which are limited to polar materials, strain skyrmions operate at room temperature with negligible dissipation and are universal to all twisted van der Waals bilayers with $C_n$ symmetry\cite{mesple2023giant}. 

Furthermore, in sliding ferroelectric moir\'e systems, lattice relaxation could be considered as electric dipoles and facilitates the modulation of the moir\'e potential energy, thereby controlling the moir\'e ferroelectric domains and regulating their thermal stability\cite{li2017binary,zheng2020unconventional}, thus this demonstrates that strain defects specifically via interlayer sliding—can serve as a powerful, purely mechanical control knob for manipulating skyrmions across both van der Waals heterostructures and sliding ferroelectric systems.

\section{Conclusion}
In summary, utilizing a continuum elastic model system under the process of adiabatic evolution, we have systematically investigated the relaxation of in-plane and out-of-plane structural displacement fields in twisted bilayer graphene. We demonstrate that this mechanical relaxation naturally gives rise to stable, three-dimensional strain skyrmions in real space. Furthermore, by introducing continuous interlayer sliding, we discover that these skyrmions exhibit a highly controllable, transverse linear Hall effect. By generalizing our model to encompass various 2D chiral point group symmetries, we have proven the universality of this purely mechanical topological pumping mechanism. We have obtained consistent conclusions across different chiral twisted systems. These findings provide a robust theoretical foundation for the mechanical manipulation of topological quasiparticles, potentially opening new avenues for the design of novel chiral material devices and high-speed, mechanically driven information transport architectures. 

\section{Acknowledgments}
We thank Yang Long for stimulating discussions. This work is supported by the National Natural Science Fundation of China (Grant No. 12404279), the Quantum Science and Technology-National Science and Technology Major Project (Grant No. 2023ZD0300500), and the National Key R\&D Program of China (Grant Nos. 2023YFA1406900 and 2022YFA1404400).

\bibliography{main}

@article{nam2017lattice,
  title = {Lattice relaxation and energy band modulation in twisted bilayer graphene},
  author = {Nam, Nguyen N. T. and Koshino, Mikito},
  journal = {Phys. Rev. B},
  volume = {96},
  issue = {7},
  pages = {075311},
  numpages = {12},
  year = {2017},
  month = {Aug},
  publisher = {American Physical Society},
  doi = {10.1103/PhysRevB.96.075311},
  url = {https://link.aps.org/doi/10.1103/PhysRevB.96.075311}
}

@article{koshino2019moir,
  title = {Moir\'e phonons in twisted bilayer graphene},
  author = {Koshino, Mikito and Son, Young-Woo},
  journal = {Phys. Rev. B},
  volume = {100},
  issue = {7},
  pages = {075416},
  numpages = {10},
  year = {2019},
  month = {Aug},
  publisher = {American Physical Society},
  doi = {10.1103/PhysRevB.100.075416},
  url = {https://link.aps.org/doi/10.1103/PhysRevB.100.075416}
}

@article{suri2021chiral,
title = {Chiral Phonons in Moiré Superlattices},
author = {Suri, Nishchay and Wang, Chong and Zhang, Yinhan and Xiao, Di},
journal = {Nano Lett.},
volume = {21},
number = {23},
pages = {10026-10031},
year = {2021},
doi = {10.1021/acs.nanolett.1c03692},
URL = {https://doi.org/10.1021/acs.nanolett.1c03692    }
}

@article{lin2018shear,
  title = {Shear instability in twisted bilayer graphene},
  author = {Lin, Xianqing and Liu, Dan and Tom\'anek, David},
  journal = {Phys. Rev. B},
  volume = {98},
  issue = {19},
  pages = {195432},
  numpages = {8},
  year = {2018},
  month = {Nov},
  publisher = {American Physical Society},
  doi = {10.1103/PhysRevB.98.195432},
  url = {https://link.aps.org/doi/10.1103/PhysRevB.98.195432}
}

@article{li2017binary,
title = {Binary Compound Bilayer and Multilayer with Vertical Polarizations: Two-Dimensional Ferroelectrics, Multiferroics, and Nanogenerators},
author = {Li, Lei and Wu, Menghao},
journal = {ACS Nano},
volume = {11},
number = {6},
pages = {6382-6388},
year = {2017},
doi = {10.1021/acsnano.7b02756},
URL = { https://doi.org/10.1021/acsnano.7b02756}
}

@article{zheng2020unconventional,
  title={Unconventional ferroelectricity in moir{\'e} heterostructures},
  author={Zheng, Zhiren and Ma, Qiong and Bi, Zhen and de La Barrera, Sergio and Liu, Ming-Hao and Mao, Nannan and Zhang, Yang and Kiper, Natasha and Watanabe, Kenji and Taniguchi, Takashi and others},
  journal={Nature},
  volume={588},
  number={7836},
  pages={71--76},
  year={2020},
  publisher={Nature Publishing Group UK London},
  doi={10.1038/s41586-020-2970-9},
  URL = {https://doi.org/10.1038/s41586-020-2970-9}
}

@article{hou2025strain,
author = {Hou, Yuan and Zhou, Jingzhuo and Xue, Minmin and Yu, Maolin and Han, Ying and Zhang, Zhuhua and Lu, Yang},
title = {Strain Engineering of Twisted Bilayer Graphene: The Rise of Strain-Twistronics},
journal = {Small},
volume = {21},
number = {28},
pages = {2311185},
doi = {https://doi.org/10.1002/smll.202311185},
url = {https://onlinelibrary.wiley.com/doi/abs/10.1002/smll.202311185},
year = {2025}
}

@article{koren2016moire,
  title = {Moir\'e scaling of the sliding force in twisted bilayer graphene},
  author = {Koren, E. and Duerig, U.},
  journal = {Phys. Rev. B},
  volume = {94},
  issue = {4},
  pages = {045401},
  numpages = {11},
  year = {2016},
  month = {Jul},
  publisher = {American Physical Society},
  doi = {10.1103/PhysRevB.94.045401},
  url = {https://link.aps.org/doi/10.1103/PhysRevB.94.045401}
}

@article{PhysRevMaterials.9.024002,
  title = {Restriction of macroscopic structural superlubricity due to structure relaxation by the example of twisted graphene bilayer},
  author = {Minkin, Alexander S. and Lebedeva, Irina V. and Popov, Andrey M. and Vyrko, Sergey A. and Poklonski, Nikolai A. and Lozovik, Yurii E.},
  journal = {Phys. Rev. Mater.},
  volume = {9},
  issue = {2},
  pages = {024002},
  numpages = {11},
  year = {2025},
  month = {Feb},
  publisher = {American Physical Society},
  doi = {10.1103/PhysRevMaterials.9.024002},
  url = {https://link.aps.org/doi/10.1103/PhysRevMaterials.9.024002}
}

@article{thouless1983quantization,
  title = {Quantization of particle transport},
  author = {Thouless, D. J.},
  journal = {Phys. Rev. B},
  volume = {27},
  issue = {10},
  pages = {6083--6087},
  numpages = {0},
  year = {1983},
  month = {May},
  publisher = {American Physical Society},
  doi = {10.1103/PhysRevB.27.6083},
  url = {https://link.aps.org/doi/10.1103/PhysRevB.27.6083}
}

@article{zhang2020topological,
  title = {Topological charge pumping in twisted bilayer graphene},
  author = {Zhang, Yinhan and Gao, Yang and Xiao, Di},
  journal = {Phys. Rev. B},
  volume = {101},
  issue = {4},
  pages = {041410(R)},
  numpages = {5},
  year = {2020},
  month = {Jan},
  publisher = {American Physical Society},
  doi = {10.1103/PhysRevB.101.041410},
  url = {https://link.aps.org/doi/10.1103/PhysRevB.101.041410}
}

@article{sarkar2005evidence,
  title={Evidence for strain glass in the ferroelastic-martensitic system Ti 50-x Ni 50+ x},
  author={Sarkar, Shampa and Ren, Xiaobing and Otsuka, Kazuhiro},
  journal={Phys. Rev. Lett.},
  volume={95},
  number={20},
  pages={205702},
  year={2005},
  publisher={APS},
   doi = {10.1103/PhysRevLett.95.205702},
  url = {https://link.aps.org/doi/10.1103/PhysRevLett.95.205702}
}

@article{wang2010strain,
  title={Strain glass in Fe-doped Ti--Ni},
  author={Wang, Dong and Zhang, Zhen and Zhang, Jian and Zhou, Yumei and Wang, Yu and Ding, Xiangdong and Wang, Yunzhi and Ren, Xiaobing},
  journal={Acta Mater.},
  volume={58},
  number={18},
  pages={6206--6215},
  year={2010},
  publisher={Elsevier},
  doi = {https://doi.org/10.1016/j.actamat.2010.07.040},
url = {https://www.sciencedirect.com/science/article/pii/S1359645410004842}
}

@article{wang2006shape,
  title={Shape memory effect and superelasticity in a strain glass alloy},
  author={Wang, Yu and Ren, Xiaobing and Otsuka, Kazuhiro},
  journal={Phys. Rev. Lett.},
  volume={97},
  number={22},
  pages={225703},
  year={2006},
  publisher={APS},
  doi={10.1103/PhysRevLett.97.225703},
  URL = {https://doi.org/10.1103/PhysRevLett.97.225703}
}

@article{xu2024polymer,
  title={A polymer-like ultrahigh-strength metal alloy},
  author={Xu, Zhizhi and Ji, Yuanchao and Liu, Chang and He, Liqiang and Zhao, Hui and Yuan, Ye and Qian, Yu and Cui, Jin and Xiao, Andong and Wang, Wenjia and others},
  journal={Nature},
  volume={633},
  number={8030},
  pages={575--581},
  year={2024},
  publisher={Nature Publishing Group UK London},
  doi={10.1038/s41586-024-07900-4},
  URL={https://doi.org/10.1038/s41586-024-07900-4}
}

@article{liu2009large,
  title = {Large Piezoelectric Effect in Pb-Free Ceramics},
  author = {Liu, Wenfeng and Ren, Xiaobing},
  journal = {Phys. Rev. Lett.},
  volume = {103},
  issue = {25},
  pages = {257602},
  numpages = {4},
  year = {2009},
  month = {Dec},
  publisher = {American Physical Society},
  doi = {10.1103/PhysRevLett.103.257602},
  url = {https://link.aps.org/doi/10.1103/PhysRevLett.103.257602}
}

@article{ren2004large,
  title={Large electric-field-induced strain in ferroelectric crystals by point-defect-mediated reversible domain switching},
  author={Ren, Xiaobing},
  journal={Nat. Mater.},
  volume={3},
  number={2},
  pages={91--94},
  year={2004},
  publisher={Nature Publishing Group UK London},
  doi={https://doi.org/10.1038/nmat1051},
URL={https://doi.org/10.1038/nmat1051}}

@article{zhang2016accelerating,
  title={Accelerating ferroic ageing dynamics upon cooling},
  author={Zhang, Junyan and Mao, Yunwei and Wang, Dong and Li, Ju and Wang, Yunzhi},
  journal={NPG Asia Mater.},
  volume={8},
  number={10},
  pages={e319--e319},
  year={2016},
  publisher={Nature Publishing Group},
  doi={10.1038/am.2016.152},
  URL={https://doi.org/10.1038/am.2016.152}
}

@article{dai2019strain,
author = {Dai, Zhaohe and Liu, Luqi and Zhang, Zhong},
title = {Strain Engineering of 2D Materials: Issues and Opportunities at the Interface},
journal = {Adv. Mater.},
volume = {31},
number = {45},
year = {2019},
pages = {1805417},
doi = {https://doi.org/10.1002/adma.201805417},
url = {https://advanced.onlinelibrary.wiley.com/doi/abs/10.1002/adma.201805417}
}

@article{peng2020strain,
  title={Strain engineering of 2D semiconductors and graphene: from strain fields to band-structure tuning and photonic applications},
  author={Peng, Zhiwei and Chen, Xiaolin and Fan, Yulong and Srolovitz, David J and Lei, Dangyuan},
  journal={Light Sci. Appl.},
  volume={9},
  number={1},
  pages={190},
  year={2020},
  publisher={Nature Publishing Group UK London},
  doi = {10.1038/s41377-020-00421-5},
  URL = {https://doi.org/10.1038/s41377-020-00421-5}
}

@article{feng2012strain,
  title={Strain-engineered artificial atom as a broad-spectrum solar energy funnel},
  author={Feng, Ji and Qian, Xiaofeng and Huang, Cheng-Wei and Li, Ju},
  journal={Nat. Photonics},
  volume={6},
  number={12},
  pages={866--872},
  year={2012},
  publisher={Nature Publishing Group UK London},
  doi={10.1038/nphoton.2012.285},
  URL={https://doi.org/10.1038/nphoton.2012.285}

}

@article{si2016strain,
  title={Strain engineering of graphene: a review},
  author={Si, Chen and Sun, Zhimei and Liu, Feng},
  journal={Nanoscale},
  volume={8},
  number={6},
  pages={3207--3217},
  year={2016},
  publisher={Royal Society of Chemistry},
  doi = {10.1039/C5NR07755A},
  URL = {https://doi.org/10.1039/C5NR07755A}
}

@article{koren2015adhesion,
author = {Elad Koren  and Emanuel Lörtscher  and Colin Rawlings  and Armin W. Knoll  and Urs Duerig },
title = {Adhesion and friction in mesoscopic graphite contacts},
journal = {Science},
volume = {348},
number = {6235},
pages = {679-683},
year = {2015},
doi = {10.1126/science.aaa4157},
URL = {https://www.science.org/doi/abs/10.1126/science.aaa4157}}

@article{nika2017phonons,
doi = {10.1088/1361-6633/80/3/036502},
url = {https://doi.org/10.1088/1361-6633/80/3/036502},
year = {2017},
month = {jan},
publisher = {IOP Publishing},
volume = {80},
number = {3},
pages = {036502},
author = {Nika, Denis L and Balandin, Alexander A},
title = {Phonons and thermal transport in graphene and graphene-based materials},
journal = {Rep. Prog. Phys.}
}

@article{bonini2012acoustic,
author = {Bonini, Nicola and Garg, Jivtesh and Marzari, Nicola},
title = {Acoustic Phonon Lifetimes and Thermal Transport in Free-Standing and Strained Graphene},
journal = {Nano Lett.},
volume = {12},
number = {6},
pages = {2673-2678},
year = {2012},
doi = {10.1021/nl202694m},
URL = {  https://doi.org/10.1021/nl202694m}
}

@article{chen2012thermal,
  title={Thermal transport in graphene supported on copper},
  author={Chen, Liang and Kumar, Satish},
  journal={J. Appl. Phys.},
  volume={112},
  number={4},
  year={2012},
  publisher={AIP Publishing},
      doi = {10.1063/1.4740071},
    url = {https://doi.org/10.1063/1.4740071}
}

@article{jiang2018manipulation,
  title={Manipulation of domain-wall solitons in bi-and trilayer graphene},
  author={Jiang, Lili and Wang, Sheng and Shi, Zhiwen and Jin, Chenhao and Utama, M Iqbal Bakti and Zhao, Sihan and Shen, Yuen-Ron and Gao, Hong-Jun and Zhang, Guangyu and Wang, Feng},
  journal={Nat. Nanotechnol.},
  volume={13},
  number={3},
  pages={204--208},
  year={2018},
  publisher={Nature Publishing Group UK London},
  doi={10.1038/s41565-017-0042-6},
  URL={https://doi.org/10.1038/s41565-017-0042-6}
}

@article{yang2020tunable,
  title={Tunable second harmonic generation in twisted bilayer graphene},
  author={Yang, Fuyi and Song, Wenshen and Meng, Fanhao and Luo, Fuchuan and Lou, Shuai and Lin, Shuren and Gong, Zilun and Cao, Jinhua and Barnard, Edward S and Chan, Emory and others},
  journal={Matter.},
  volume={3},
  number={4},
  pages={1361--1376},
  year={2020},
  publisher={Elsevier},
  doi={10.1016/j.matt.2020.08.018},
  URL={https://doi.org/10.1016/j.matt.2020.08.018}
}

@article{mesple2023giant,
author = {Li, Lei and Wu, Menghao},
title = {Binary Compound Bilayer and Multilayer with Vertical Polarizations: Two-Dimensional Ferroelectrics, Multiferroics, and Nanogenerators},
journal = {ACS Nano},
volume = {11},
number = {6},
pages = {6382-6388},
year = {2017},
doi = {10.1021/acsnano.7b02756},
URL = {        https://doi.org/10.1021/acsnano.7b02756}
}

@article{wang2024thermodynamic,
  title = {Thermodynamic Geometry of Nonequilibrium Fluctuations in Cyclically Driven Transport},
  author = {Wang, Zi and Ren, Jie},
  journal = {Phys. Rev. Lett.},
  volume = {132},
  issue = {20},
  pages = {207101},
  numpages = {8},
  year = {2024},
  month = {May},
  publisher = {American Physical Society},
  doi = {10.1103/PhysRevLett.132.207101},
  url = {https://link.aps.org/doi/10.1103/PhysRevLett.132.207101}
}

@article{lei2025quantum,
author = {Lei, Changyong and Wang, Zi and Yang, Chenwen and Liu, Yizhou and Ren, Jie},
title = {Quantum Phononics: From Principles to Engineering},
journal = {J. Phys. Chem. Lett.},
volume = {16},
number = {30},
pages = {7630-7641},
year = {2025},
doi = {10.1021/acs.jpclett.5c00951},
URL = { https://doi.org/10.1021/acs.jpclett.5c00951
 }
}

@article{
long2018intrinsic,
author = {Yang Long  and Jie Ren  and Hong Chen },
title = {Intrinsic spin of elastic waves},
journal = {Proc. Natl. Acad. Sci.},
volume = {115},
number = {40},
pages = {9951-9955},
year = {2018},
doi = {10.1073/pnas.1808534115},
URL = {https://www.pnas.org/doi/abs/10.1073/pnas.1808534115}
}

@article{shi2019observation,
  title={Observation of acoustic spin},
  author={Shi, Chengzhi and Zhao, Rongkuo and Long, Yang and Yang, Sui and Wang, Yuan and Chen, Hong and Ren, Jie and Zhang, Xiang},
  journal={Natl. Sci. Rev.},
  volume={6},
  number={4},
  pages={707--712},
  year={2019},
  publisher={Oxford University Press},
      doi = {10.1093/nsr/nwz059},
    url = {https://doi.org/10.1093/nsr/nwz059}
}

@article{long2020realization,
  title={Realization of acoustic spin transport in metasurface waveguides},
  author={Long, Yang and Zhang, Danmei and Yang, Chenwen and Ge, Jianmin and Chen, Hong and Ren, Jie},
  journal={Nat. Commun.},
  volume={11},
  number={1},
  pages={4716},
  year={2020},
  publisher={Nature Publishing Group UK London},
  doi={10.1038/s41467-020-18599-y},
  URL={https://doi.org/10.1038/s41467-020-18599-y}
}

@article{long2020symmetry,
  title={Symmetry selective directionality in near-field acoustics},
  author={Long, Yang and Ge, Hao and Zhang, Danmei and Xu, Xiangyuan and Ren, Jie and Lu, Ming-Hui and Bao, Ming and Chen, Hong and Chen, Yan-Feng},
  journal={Natl. Sci. Rev.},
  volume={7},
  number={6},
  pages={1024--1035},
  year={2020},
  publisher={Oxford University Press},
     doi = {10.1093/nsr/nwaa040},
    url = {https://doi.org/10.1093/nsr/nwaa040}
}

@article{yuan2021observation,
  title={Observation of elastic spin with chiral meta-sources},
  author={Yuan, Weitao and Yang, Chenwen and Zhang, Danmei and Long, Yang and Pan, Yongdong and Zhong, Zheng and Chen, Hong and Zhao, Jinfeng and Ren, Jie},
  journal={Nat. Commun.},
  volume={12},
  number={1},
  pages={6954},
  year={2021},
  publisher={Nature Publishing Group UK London},
  doi={10.1038/s41467-021-27254-z},
  URL={https://doi.org/10.1038/s41467-021-27254-z}
}

@article{zhao2022elastic,
  title={Elastic valley spin controlled chiral coupling in topological valley phononic crystals},
  author={Zhao, Jinfeng and Yang, Chenwen and Yuan, Weitao and Zhang, Danmei and Long, Yang and Pan, Yongdong and Chen, Hong and Zhong, Zheng and Ren, Jie},
  journal={Phys. Rev. Lett.},
  volume={129},
  number={27},
  pages={275501},
  year={2022},
  publisher={APS},
    doi = {10.1103/PhysRevLett.129.275501},
  url = {https://link.aps.org/doi/10.1103/PhysRevLett.129.275501}
}

@article{ren2022elastic,
doi = {10.1088/0256-307X/39/12/126301},
url = {https://doi.org/10.1088/0256-307X/39/12/126301},
year = {2022},
month = {dec},
publisher = {Chinese Physical Society and IOP Publishing Ltd},
volume = {39},
number = {12},
pages = {126301},
author = {Ren, Jie},
title = {From Elastic Spin to Phonon Spin: Symmetry and Fundamental Relations},
journal = {Chinese Phys. Lett.},
}

@article{yang2023hybrid,
  title = {Hybrid Spin and Anomalous Spin-Momentum Locking in Surface Elastic Waves},
  author = {Yang, Chenwen and Zhang, Danmei and Zhao, Jinfeng and Gao, Wenting and Yuan, Weitao and Long, Yang and Pan, Yongdong and Chen, Hong and Nori, Franco and Bliokh, Konstantin Y. and Zhong, Zheng and Ren, Jie},
  journal = {Phys. Rev. Lett.},
  volume = {131},
  issue = {13},
  pages = {136102},
  numpages = {6},
  year = {2023},
  month = {Sep},
  publisher = {American Physical Society},
  doi = {10.1103/PhysRevLett.131.136102},
  url = {https://link.aps.org/doi/10.1103/PhysRevLett.131.136102}
}

@article{
yang2024chirality,
author = {Chenwen Yang  and Jie Ren },
title = {Chirality-induced phonon spin selectivity by elastic spin–orbit interaction},
journal = {Proc. Natl. Acad. Sci.},
volume = {121},
number = {47},
pages = {e2411427121},
year = {2024},
doi = {10.1073/pnas.2411427121},
URL = {https://www.pnas.org/doi/abs/10.1073/pnas.2411427121}}

@article{liu2021chirality,
  title={Chirality-driven topological electronic structure of DNA-like materials},
  author={Liu, Yizhou and Xiao, Jiewen and Koo, Jahyun and Yan, Binghai},
  journal={Nat. Mater.},
  volume={20},
  number={5},
  pages={638--644},
  year={2021},
  publisher={Nature Publishing Group UK London},
  doi={10.1038/s41563-021-00924-5},
  URL={https://doi.org/10.1038/s41563-021-00924-5}
}

@article{liu2017pseudospins,
  title = {Pseudospins and Topological Effects of Phonons in a Kekul\'e Lattice},
  author = {Liu, Yizhou and Lian, Chao-Sheng and Li, Yang and Xu, Yong and Duan, Wenhui},
  journal = {Phys. Rev. Lett.},
  volume = {119},
  issue = {25},
  pages = {255901},
  numpages = {6},
  year = {2017},
  month = {Dec},
  publisher = {American Physical Society},
  doi = {10.1103/PhysRevLett.119.255901},
  url = {https://link.aps.org/doi/10.1103/PhysRevLett.119.255901}
}

@article{liu2018berry,
    author = {Liu, Yizhou and Xu, Yong and Duan, Wenhui},
    title = {Berry phase and topological effects of phonons},
    journal = {Nat. Sci. Rev.},
    volume = {5},
    number = {3},
    pages = {314-316},
    year = {2018},
    month = {05},
    issn = {2095-5138},
    doi = {10.1093/nsr/nwx086},
    url = {https://doi.org/10.1093/nsr/nwx086}
}

@article{liu2020topological,
author = {Liu, Yizhou and Chen, Xiaobin and Xu, Yong},
title = {Topological Phononics: From Fundamental Models to Real Materials},
journal = {Adv. Funct. Mater.},
volume = {30},
number = {8},
pages = {1904784},
year = {2020},
doi = {https://doi.org/10.1002/adfm.201904784},
url = {https://advanced.onlinelibrary.wiley.com/doi/abs/10.1002/adfm.201904784}
}

@article{liu2022ubiquitous,
  title={Ubiquitous topological states of phonons in solids: Silicon as a model material},
  author={Liu, Yizhou and Zou, Nianlong and Zhao, Sibo and Chen, Xiaobin and Xu, Yong and Duan, Wenhui},
  journal={Nano Lett.},
  volume={22},
  number={5},
  pages={2120--2126},
  year={2022},
  publisher={ACS Publications}
}

@article{PhysRevLett.108.086602,
  title = {Phonon-Induced Backscattering in Helical Edge States},
  author = {Budich, Jan Carl and Dolcini, Fabrizio and Recher, Patrik and Trauzettel, Bj\"orn},
  journal = {Phys. Rev. Lett.},
  volume = {108},
  issue = {8},
  pages = {086602},
  numpages = {4},
  year = {2012},
  month = {Feb},
  publisher = {American Physical Society},
  doi = {10.1103/PhysRevLett.108.086602},
  url = {https://link.aps.org/doi/10.1103/PhysRevLett.108.086602}
}

@misc{my_supplementary,
  note  = {Supporting Information.}
}

@article{wei2012nature,
  title={The nature of strength enhancement and weakening by pentagon--heptagon defects in graphene},
  author={Wei, Yujie and Wu, Jiangtao and Yin, Hanqing and Shi, Xinghua and Yang, Ronggui and Dresselhaus, Mildred},
  journal={Nature materials},
  volume={11},
  number={9},
  pages={759--763},
  year={2012},
  publisher={Nature Publishing Group UK London},
  doi={10.1038/nmat3370},
  URL={https://doi.org/10.1038/nmat3370}
}

@article{yazyev2010topological,
  title={Topological defects in graphene: Dislocations and grain boundaries},
  author={Yazyev, Oleg V and Louie, Steven G},
  journal={Physical Review B—Condensed Matter and Materials Physics},
  volume={81},
  number={19},
  pages={195420},
  year={2010},
  publisher={APS},
  doi={10.1103/PhysRevB.81.195420},
  URL={ https://doi.org/10.1103/PhysRevB.81.195420}
}

@article{RevModPhys.81.109,
  title = {The electronic properties of graphene},
  author = {Castro Neto, A. H. and Guinea, F. and Peres, N. M. R. and Novoselov, K. S. and Geim, A. K.},
  journal = {Rev. Mod. Phys.},
  volume = {81},
  issue = {1},
  pages = {109--162},
  numpages = {0},
  year = {2009},
  month = {Jan},
  publisher = {American Physical Society},
  doi = {10.1103/RevModPhys.81.109},
  url = {https://link.aps.org/doi/10.1103/RevModPhys.81.109}
}

@book{nelson2002defects,
  title={Defects and geometry in condensed matter physics},
  author={Nelson, David R},
  year={2002},
  publisher={Cambridge University Press},
  url={https://books.google.co.jp/books?id=YtYFAqswRzUC&lpg=PR11&ots=0eMdKOuxnv&dq=D.%20R.%20Nelson%2C%20Defects%20and%20Geometry%20in%20Condensed%20Matter%20Phys%20ics%20Cambridge%20University%20Press%2C%20Cambridge%2C%202002.&lr&pg=PP1#v=onepage&q=D.%20R.%20Nelson,%20Defects%20and%20Geometry%20in%20Condensed%20Matter%20Phys%20ics%20Cambridge%20University%20Press,%20Cambridge,%202002.&f=false}
}

@article{RevModPhys.86.253,
  title = {First-principles calculations for point defects in solids},
  author = {Freysoldt, Christoph and Grabowski, Blazej and Hickel, Tilmann and Neugebauer, J\"org and Kresse, Georg and Janotti, Anderson and Van de Walle, Chris G.},
  journal = {Rev. Mod. Phys.},
  volume = {86},
  issue = {1},
  pages = {253--305},
  numpages = {53},
  year = {2014},
  month = {Mar},
  publisher = {American Physical Society},
  doi = {10.1103/RevModPhys.86.253},
  url = {https://link.aps.org/doi/10.1103/RevModPhys.86.253}
}

@article{Macdonald_1992,
doi = {10.1149/1.2069096},
url = {https://doi.org/10.1149/1.2069096},
year = {1992},
month = {dec},
publisher = {The Electrochemical Society, Inc.},
volume = {139},
number = {12},
pages = {3434},
author = {Macdonald, Digby D.},
title = {The Point Defect Model for the Passive State},
journal = {Journal of The Electrochemical Society},
}

@article{MACDONALD20111761,
title = {The history of the Point Defect Model for the passive state: A brief review of film growth aspects},
journal = {Electrochimica Acta},
volume = {56},
number = {4},
pages = {1761-1772},
year = {2011},
note = {ADVANCES IN CORROSION SCIENCE FOR LIFETIME PREDICTION AND SUSTAINABILITYSelection of papers from the 8th ISE Spring Meeting2-5 May 2010, Columbus, OH, USA},
issn = {0013-4686},
doi = {https://doi.org/10.1016/j.electacta.2010.11.005},
url = {https://www.sciencedirect.com/science/article/pii/S001346861001515X}}

@article{Stukowski_2010,
doi = {10.1088/0965-0393/18/8/085001},
url = {https://doi.org/10.1088/0965-0393/18/8/085001},
year = {2010},
month = {sep},
publisher = {},
volume = {18},
number = {8},
pages = {085001},
author = {Stukowski, Alexander and Albe, Karsten},
title = {Extracting dislocations and non-dislocation crystal defects from atomistic simulation data},
journal = {Modelling and Simulation in Materials Science and Engineering}
}

@book{hull2011introduction,
  title={Introduction to dislocations},
  author={Hull, Derek and Bacon, David J},
  volume={37},
  year={2011},
  publisher={Elsevier},
  url={https://books.google.co.jp/books?id=MvSvqll6ct0C&lpg=PP1&ots=DaHf_mz_P-&dq=dislocations%20defect&lr&pg=PP1#v=onepage&q=dislocations%20defect&f=false}
}

@article{KOMA1999236,
title = {Van der Waals epitaxy for highly lattice-mismatched systems},
journal = {Journal of Crystal Growth},
volume = {201-202},
pages = {236-241},
year = {1999},
issn = {0022-0248},
doi = {https://doi.org/10.1016/S0022-0248(98)01329-3},
url = {https://www.sciencedirect.com/science/article/pii/S0022024898013293},
author = {Atsushi Koma}
}

@article{doi.10.1021,
author = {Liu, Jia and Zhang, Jiatao},
title = {Nanointerface Chemistry: Lattice-Mismatch-Directed Synthesis and Application of Hybrid Nanocrystals},
journal = {Chemical Reviews},
volume = {120},
number = {4},
pages = {2123-2170},
year = {2020},
doi = {10.1021/acs.chemrev.9b00443},
    note ={PMID: 31971378},

URL = { https://doi.org/10.1021/acs.chemrev.9b00443}
}

@article{PhysRevLett.111.196802,
  title = {First-Principles Calculations on the Effect of Doping and Biaxial Tensile Strain on Electron-Phonon Coupling in Graphene},
  author = {Si, Chen and Liu, Zheng and Duan, Wenhui and Liu, Feng},
  journal = {Phys. Rev. Lett.},
  volume = {111},
  issue = {19},
  pages = {196802},
  numpages = {5},
  year = {2013},
  month = {Nov},
  publisher = {American Physical Society},
  doi = {10.1103/PhysRevLett.111.196802},
  url = {https://link.aps.org/doi/10.1103/PhysRevLett.111.196802}
}

\end{document}